\documentclass[aps,superscriptaddress,twocolumn]{revtex4}

\usepackage{amsmath}
\usepackage{graphicx}

\begin{document}

\title{Electrostatic pair creation and recombination in quantum plasmas}

\author{M.\ Marklund}
\affiliation{Centre for Nonlinear Physics, Department of Physics,
Ume{\aa} University, SE-901 87 Ume{\aa}, Sweden}

\author{B.\ Eliasson}
\affiliation{Institut f\"ur Theoretische Physik IV, Ruhr-Universit\"at
Bochum, D--44780 Bochum,
   Germany}

\author{P.\ K.\ Shukla}
\affiliation{Centre for Nonlinear Physics, Department of Physics,
Ume{\aa} University, SE-901 87 Ume{\aa}, Sweden}
\affiliation{Institut f\"ur Theoretische Physik IV, Ruhr-Universit\"at
Bochum, D--44780 Bochum,
   Germany}

\author{L. Stenflo}
\affiliation{Centre for Nonlinear Physics, Department of Physics,
Ume{\aa} University, SE-901 87 Ume{\aa}, Sweden}

\author{M. E. Dieckmann}
\affiliation{Institut f\"ur Theoretische Physik IV, Ruhr-Universit\"at
Bochum, D--44780 Bochum,
   Germany}

\author{M. Parviainen}
\affiliation{Institut f\"ur Theoretische Physik IV, Ruhr-Universit\"at
Bochum, D--44780 Bochum,
   Germany}

\begin{abstract}
   The collective production of electron--positron pairs by
electrostatic waves in quantum plasmas is investigated. In particular,
a semi-classical governing set of equation for a self-consistent
treatment of pair creation by the Schwinger mechanism in a quantum
plasma is derived.
\end{abstract}
\pacs{}

\maketitle
\section{Introduction}
In quantum field theory, and quantum electrodynamics (QED) in
particular, the vacuum
is no longer trivial, but may under certain circumstances act as a
medium. Effects
such as elastic light-by-light scattering as photon splitting are well
known examples of
the influence of the nonlinear quantum vacuum \cite{marklund-shukla}.
The concept of pair creation is of great interest both from a principal
point of view,
as well as for near-future applications. In particular, the production
of pairs in
an intrinsically nonperturbative process, and as such poses new
requirements
for its theoretical treatment. Lately, there has been much interest in
pair creation due
to nontrivial generalizations of the constant electric field case (see
e.g.
\cite{brezin-itzykson,alkofer-etal,nitta-etal,blaschke-etal,%
gies-klingmuller,gies-etal,dunne-schubert,fried-gabellini}
and references therein), as considered by Schwinger for quantum
electrodynamics
\cite{Schwinger}. The next generation laser systems, such as the X-ray
free electron
laser, will be able to produce pairs, and this necessitates the study
of collective and
self-consistent pair production effects. Thus, in this spirit, the
effects of pair
production on circularly polarized light was considered in Refs.\
\cite{bulanov} and
\cite{Bulanov-etal}, and it was found that electromagnetic degrees of
freedom were
dissipated into pairs in a self-consistent manner. However, it is not
unreasonable
to assume that within laser-plasma and astrophysical systems, strong
electrostatic
fields may be built up, thus posing a new problem concerning pair
creation.

In this paper we will study the effects of electrostatic pair creation
in a
three component plasma, consisting of ions, electrons, and positrons.
The quantum properties of electrons and positrons are partially included
by using a semi-classical approximation of the many-body Schr\"odinger
system. The effect of the latter is to introduce higher order dispersion
in the electron/positron momentum equations. The combined effect of
pair creation and quantum plasmas can thus be studied using and
effective
semi-classical theory in terms of macroscopic fluid variables.
\section{Theory}
For fields varying slowly compared to the Compton frequency $\omega_e =
m_ec^2/\hbar$, where $m_e$ is the electron rest mass, $c$ is the
velocity of light in vacuum, and $\hbar$ is the Planck constant divided
by $2\pi$, the
pair creation rate per unit phase space volume given by
$q(t,\vec{r},\vec{p}) = q_0(t,\vec{r})F(\vec{p})$, where
\cite{Schwinger,Kajantie-Matsui,Gatoff-etal,Kluger-etal}
\begin{equation}\label{eq:creation}
   q_0 =
  \frac{c}{(2\pi)^3\lambda^4}\frac{|\vec{E}|^2}{E_{\mathrm{crit}}^2}
  \exp\left( -\pi\frac{E_{\text{crit}}}{|\vec{E}|}\right) ,
\end{equation}
Here $\vec{p}$ is the electron/positron microscopic momentum, $\vec{E}$
is the electric field strength, $e$ is the magnitude of the electron
charge, $E_{\text{crit}} = m_e^2c^3/e\hbar \sim
10^{16}\,\mathrm{V\,cm^{-1}}$ is the Schwinger critical field, $\lambda
= \hbar/m_ec$ is the Compton wavelength, $m_e$ is the electron rest
mass, $c$ is the speed of light in vacuum, $\hbar$ in Planck's constant
divided by $2\pi$, and the momentum distribution function is normalized
according to $\int\,Fd^3p = 1$. 
produced at rest, thus the delta function in momentum space, and
neglected the small correction due to the electron/positron momentum
orthogonal to the electric field.
We note that the expression (\ref{eq:creation}) should be used with
caution, as pointed out in Ref.\ \cite{gies-klingmuller}, as the
electrons/positrons does not experience the
local field. This can be partially remedied by averaging the field over
a suitable spatial volume. Here we will at present consider the local
field approximation, keeping in mind the above. Apart from the pair
creation rate,
there will also be a pair recombination, or annihilation, rate given by
$\nu_{e} = \sigma v_c n_{p}$ ($\nu_p = \sigma v_c n_{e}$) for the
electrons (positrons),
where $\sigma$ is the cross section, $v_c$ is a characteristic velocity
of the recombination, and $n_e$ ($n_p$) is the electron (positron)
number density.

For electrostatic oscillations, the electric field is determined by
Poisson's equation
\begin{subequations}
\begin{equation}\label{eq:poisson}
   \vec{\nabla}\cdot\vec{E} = (-e n_e + e n_p + Z_i e n_i+\rho_{\rm
pol})/\epsilon_0 .
\end{equation}
where $n_{e}$ ($n_p$) is the electron (positron) number density in the
laboratory frame, $n_i$ is the ion number density,  $Z_ie$ is the ion
charge density and $\rho_{\rm pol}$ is a vacuum polarization charge
density. Amp\`ere's law gives
\begin{equation}\label{eq:ampere}
   \partial_t \vec{E}=\frac{1}{\epsilon_0}\left(\frac{e n_e
   \vec{p}_e}{\gamma_e m_e}
     - \frac{e n_p\vec{p}_p}{\gamma_p m_p}-\frac{Z_i e n_i p_i}{\gamma_i m_i} - \vec{j}_{\rm
     pol}\right),
\end{equation}
where the polarization current is $\vec{j}_{\rm
pol}=\vec{E}q_0/|\vec{E}|^2$ and the vacuum polarization
charge density is defined via the continuity equation
$\partial_t\rho_{\rm pol} + \vec{\nabla}\cdot\vec{j_{\rm pol}}=0$.
The particular choice of $\vec{j}_{\rm pol}$ gives overall energy
conservation to the system where the
energy of newly created pairs is compensated by a decrease of the
electrostatic energy.
The fluid equation for the cold electrons, positrons, and ions are
governed by \cite{Manfredi,Bulanov-etal}
\begin{eqnarray}
   &&
   \partial_t n_e + \vec{\nabla}\cdot\left(\frac{n_e \vec{p}_e}{\gamma_e m_e}\right)
     = q_0 - \sigma v_cn_en_p,
   \label{eq:electronnumber} \\
   &&
   \partial_t n_p + \vec{\nabla}\cdot\left(\frac{n_p \vec{p}_p}{\gamma_p m_p}\right)
     = q_0 - \sigma v_cn_en_p ,
   \label{eq:positronnumber} \\
   && \partial_t n_i + \vec{\nabla}\cdot\left(\frac{n_i \vec{p}_i}{\gamma_i m_i}\right) = 0 ,
   \label{eq:ionnumber}
\end{eqnarray}
and
\begin{eqnarray}
  && \!\!\!\!\! \!\!\!\!\!
   \left(\partial_t +
\frac{\vec{p}_{e}}{\gamma_e m_e}\cdot\vec{\nabla}\right)\vec{p}_e =
-e\vec{E}
   + \frac{\hbar^2}{2m_e \gamma_e}\vec{\nabla}U_{\mathrm{B}e}
   \nonumber \\
   && \qquad + \frac{q_0}{n_e}(\vec{P}_e-\vec{p}_e) - \sigma v_c n_p (\vec{p}_e - \vec{p}_p),
   \label{eq:electronmom} \\
  && \!\!\!\!\! \!\!\!\!\!
   \left(\partial_t + \frac{\vec{p}_p}{\gamma_p m_p}\cdot\vec{\nabla}\right)\vec{p}_p
= e\vec{E}
   + \frac{\hbar^2}{2m_p\gamma_p}\vec{\nabla}U_{\mathrm{B}p}
   \nonumber \\ && \qquad + 
   \frac{q_0}{n_p}(\vec{P}_p-\vec{p}_p) - \sigma v_cn_e(\vec{p}_p - \vec{p}_e),
   \label{eq:positronmom} \\
   && \!\!\!\!\! \!\!\!\!\!
   \left(\partial_t + \frac{\vec{p}_i}{\gamma_i m_i}\cdot\vec{\nabla}\right)\vec{p}_i
= Z_i e\vec{E},
   \label{eq:ionmom}
\end{eqnarray}
\label{eq:system}
\end{subequations}
valid for length scales larger than $v_F/\omega_{\mathrm{p}e}$,
$v_F$ being the Fermi velocity and $\omega_{\mathrm{p}e} =
(n_0e^2/\epsilon_0 m_e)^{1/2}$ is the electron plasma frequency for
some typical electron density $n_0$. Here we have added the
recombination term for the sake of generality, and the index $e$
($p$) denotes the electrons (positrons), $i$ denotes the ions,
$n_{e,p,i}$ is the number density, $\vec{p}_{e,p,i}$ is the fluid
momentum, $-e$ ($+e$) is the electron (positron) charge, $Z_ie$ is
the ion charge, and $\gamma_{e,p,i} = (1 + p_{e,p,i}^2/m_{e,p,i}^2
c^2)^{1/2}$ is the relativistic gamma factor. Moreover, the
generalized relativistic Bohm potential is defined according to
(neglecting the spin of the particles)
\begin{eqnarray}
   &&
   U_{\mathrm{B}e,p} =
(n_{e,p}/\gamma_{e,p})^{-1/2}\nabla^2(n_{e,p}/\gamma_{e,p})^{1/2}
   \nonumber \\ && \qquad
     -
(n_{e,p}/\gamma_{e,p}c^2)^{-1/2}\partial_t^2(n_{e,p}/\gamma_{e,p})^{1/2} ,
\label{eq:bohm}
\end{eqnarray}
obtained using a Madelung representation for the Klein-Gordon equation,
and we have defined
$\vec{P}_{e,p} = \int\,\vec{p}_{e,p}F\,d^3p$.
Equation (\ref{eq:poisson})--(\ref{eq:ionmom}), together with Eq.\
(\ref{eq:bohm}), determines the semi-classical dynamics of the three
component plasma where quantum mechanical correction for the electrons
and positrons are taken into account.

\section{One-dimensional model}
We now specialize to a one-dimensional geometry. In addition to
the effects discussed in the 3D model, we will assume that the
electron-positron pairs are created at some distance from each other 
so that the energy of the particles are compensated
by a decrease of the electric field energy. 
The Poisson equation is on the form
\begin{equation}
  \partial_x E=\frac{e}{\epsilon_0}(n_p - n_e - Z_i n_i)
  \label{Poisson_1D}
\end{equation}
where ions are taken to be immobile so that $Z_in_i=n_0$.
We assume that the the
electron and positron continuity equations, respectively, are on the form
\begin{equation}
   \partial_t n_e+\partial_x \left(\frac{n_e p_e}{\gamma_e m_e}\right)=Q_e,
\end{equation}
and
\begin{equation}
   \partial_t n_p+\partial_x\left(\frac{n_p p_p}{\gamma_p m_p}\right)=Q_p,
\end{equation}
and the momentum equations
\begin{equation}
   \partial_t p_e+m_e c^2\partial_x \gamma_e=-eE
\end{equation}
and
\begin{equation}
   \partial_t p_p+m_p c^2\partial_x \gamma_p=eE,
   \label{pp_1D}
\end{equation}
where we have assumed that the cold particles are created with momenta
equal to the
fluid momenta of the electrons and positrons, respectively, so that the
sources do not
contribute to the momenta of the particles but only to the particle
number densities.
In this manner, we do not have heating of the particles and do not need
to include pressure or take into account kinetic effects.
The source terms are taken to be on the form
\begin{equation}
   Q_{e}=q_0+\partial_x \left( \lambda \gamma_e
q_0\frac{E_{\rm crit}}{E} \right)
\end{equation}
and
\begin{equation}
   Q_{p}=q_0-\partial_x \left(\lambda \gamma_p q_0
\frac{E_{\rm crit}}{E} \right)
\end{equation}
and correspond roughly to that the creation of electrons and positrons
at the position $x$
depends on the electric field at a distance $\lambda\gamma_p E_{\rm
crit}/E$
and $-\lambda\gamma_e E_{\rm crit}/E$, respectively, from $x$. For a
stronger electric field, the
the distance of creation decreases, while it increases for larger
$\gamma_j$.
This form on the source term gives an approximate energy conservation
law
of the form
\begin{equation}
   \begin{split}
     &\frac{d}{dt}\int_{\Omega}\left[ m_e c^2 (n_e \gamma_e + n_p
\gamma_p-2n_0)+\frac{\epsilon_0 E^2}{2} \right]\,dx
   \\
     &\qquad =mc^2 \int\frac{E_{\rm crit}}{2 E}q_0\lambda\partial_x(\gamma_e^2-\gamma_p^2)\,dx,
   \end{split}
\end{equation}
where the right-hand side can be assumed to be small if
$\lambda|\partial_x|\ll 1$, and where we have assumed that all
fields and velocities vanish at $|x|=\infty$. The conservation law
was obtained by using Amp\`eres law
\begin{equation}
   \partial_t E=\frac{e}{\epsilon_0}\left[\frac{n_e
p_e}{\gamma_e m_e}
   -\frac{n_p p_p}{\gamma_p m_p}+(\gamma_e+\gamma_p)\lambda \frac{E_{\rm
crit}}{E}q_0\right],
\end{equation}
which is derived by differentiating the Poisson equation by time and using the
continuity equations for
the particles.

\section{The normalized system of equations}

We may normalize the system of equations (\ref{eq:system}) by
introducing the Compton wave length $\lambda$ and the time scale
$\tau = \hbar/m_ec^2$, normalizing the electric field by
$E_{\mathrm{crit}}$, normalizing all momenta by $m_e c$ (with
$m_p=m_e$), and normalizing the number densities by some preferred
density $n_0$. Assuming stationary ions, we then obtain the
normalized and dimensionless system
\begin{subequations}
\begin{equation}\label{eq:poisson-norm}
   \vec{\nabla}\cdot\vec{E} = \omega_{\mathrm{p}e}^2(1 - n_e + n_p) ,
\end{equation}
\begin{eqnarray}
   &&\!\!\!\!\!\!\!\!\!\!\!\!
   \partial_t n_e + \vec{\nabla}\cdot\left(\frac{n_e \vec{p}_e}{\gamma_e}\right)
     = q_0 - a n_en_p ,
   \label{eq:electronnumber-norm} \\
  && \!\!\!\!\!\!\!\!\!\!\!\!
   \partial_t n_p + \vec{\nabla}\cdot\left(\frac{n_p \vec{p}_p}{\gamma_p}\right)
     = q_0 - a n_en_p ,
   \label{eq:positronnumber-norm}
\end{eqnarray}
and
\begin{eqnarray}
  &&\!\!\!\!\!\!\!\!\!\!\!\!\!\!\!\!\!\!\!
   \left(\partial_t +
\frac{\vec{p}_{e}}{\gamma_e}\cdot\vec{\nabla}\right)\vec{p}_e =
-\vec{E}
   + \tfrac{1}{2}\vec{\nabla}U_{\mathrm{B}e}  +
\frac{q_0}{n_e}\left(\vec{P}_e - \vec{p}_e \right)
     \nonumber \\ && \qquad
     - a n_p(\vec{p}_e - \vec{p}_p),
   \label{eq:electronmom-norm} \\
  && \!\!\!\!\!\!\!\!\!\!\!\!\!\!\!\!\!\!
   \left(\partial_t + \frac{\vec{p}_p}{\gamma_p}\cdot\vec{\nabla}\right)\vec{p}_p
= \vec{E}
   + \tfrac{1}{2}\vec{\nabla}U_{\mathrm{B}p}  +
\frac{q_0}{n_p}\left(\vec{P}_p - \vec{p}_p \right)
     \nonumber \\ && \qquad
     - a n_e(\vec{p}_p - \vec{p}_e) ,
   \label{eq:positronmom-norm}
\end{eqnarray}
\label{eq:system-norm}
\end{subequations}
where now
\begin{equation}
   q_0 = ({E^2}/{N_0})\exp\left( -\pi|E|^{-1}\right)
\end{equation}
and we have used the stationary ion background density as the
normalization density  $n_0$, $N_0 = n_0h^3/m_e^3c^3$, the electron
plasma frequency is normalized by $1/\tau$ such that
$\widetilde{\omega}_{\mathrm{p}e} = (2\alpha N_0)^{1/2}/2\pi$, $\alpha =
e^2/4\pi\epsilon_0\hbar c \approx 1/137$ is the fine structure
constant, and we have denoted $a = N_0(\sigma/\lambda^2)(v_c/c)$.
\section{Numerical analysis of the one-dimensional system}

\begin{figure}
\centering
\includegraphics[width=9cm]{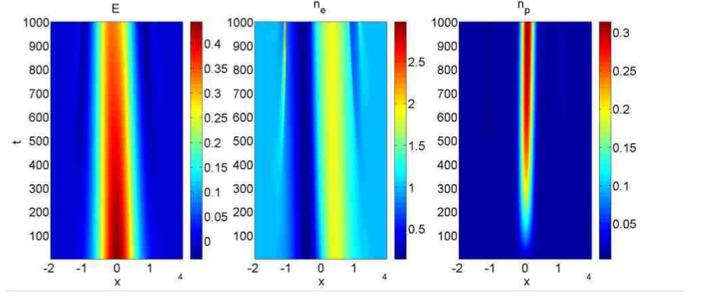}
\caption{
The electrostatic field (left panel) electron density (middle panel) and 
positron density (right panel) as a function of space $x$ and time $t$.
We used $N_0=0.2$ and the initial conditions $n_e=1.01+2(x/L)\exp(-x^2/L^2)$
with $L=6000$, and $n_p=0.01$, for the electron and positron number densities, respectively.}
\end{figure}

\begin{figure}
\centering
\includegraphics[width=8cm]{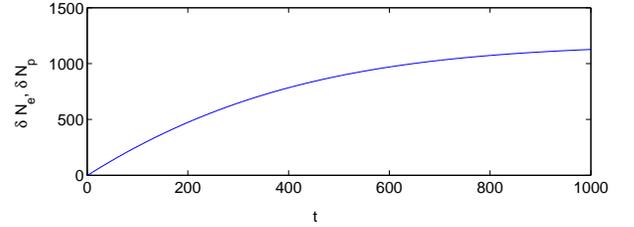}
\caption{
The total number of created electron-positron pairs, $\delta N_p=\delta N_e=N_e-N_{e, t=0}$, where
$N_e=\int_{\omega}n_e\,dx$, as a function of time. We see a decrease in the pair creation rate, correlated
with a decrease of the amplitude of the electrostatic field seen in Fig. 1.
}
\end{figure}

\begin{figure}
\centering
\includegraphics[width=8cm]{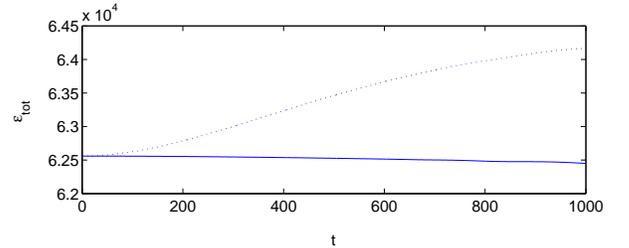}
\caption{
The total energy ${\cal E}_{tot}=
\int_{\Omega}(n_e\gamma_e+n_p\gamma_p+E^2/2\widetilde{\omega}_p^2)\,dx$
as a function of time (solid curve). The dotted curve displays the same
integral for the case when the correction for the spatial displacement 
of the created pairs, the last terms in the right-hand sides 
of Eqs. (\ref{ne_1D}) and (\ref{np_1D}), are neglected. 
In the latter case, the creation of
electron-positron pairs lead to an increase of the total energy in the system.
}
\end{figure}

The one-dimensional system of equations 
(\ref{Poisson_1D})--(\ref{pp_1D}) are normalized so that
\begin{subequations}
\begin{equation}
   \partial_x E = \widetilde{\omega}_{\mathrm{p}e}^2(1 - n_e + n_p) ,
\end{equation}
\begin{eqnarray}
   &&\!\!\!\!\!\!\!\!\!\!\!\!
   \partial_t n_e + \partial_x \left(\frac{n_e p_e}{\gamma_e}\right)
     = q_0+\partial_x\left(\frac{\gamma_e q_0}{E}\right) ,
     \label{ne_1D}
   \\
  && \!\!\!\!\!\!\!\!\!\!\!\!
   \partial_t n_p + \partial_x \left(\frac{n_p p_p}{\gamma_p}\right)
     = q_0-\partial_x\left(\frac{\gamma_e q_0}{E}\right) ,
     \label{np_1D}
\end{eqnarray}
\begin{eqnarray}
  &&\!\!\!\!\!\!\!\!\!\!\!\!\!\!\!\!\!\!\!
   \left(\partial_t + \frac{p_e}{\gamma_e}\partial_x \right)p_e =-E,
   \\
  && \!\!\!\!\!\!\!\!\!\!\!\!\!\!\!\!\!\!
   \left(\partial_t + \frac{p_p}{\gamma_p}\partial_x\right)p_p= E.
\end{eqnarray}
\end{subequations}
and are solved numerically. The results are 
displayed in Figs. (1)--(3). Initially, the electron density is perturbed
locally on the form $n_e=1.01+2(x/L)\exp(-x^2/L^2)$ with $L=6000$, 
and for the positrons
we take a small non-zero component $n_p=0.01$. All particle species are 
assumed initially to be in rest, i.e., $p_e=p_p=0$ at $t=0$. 
We have assumed a dense plasma so that $N_0=0.2$. Initially, the maximum 
electric field is approximately half the Schwinger field, 
$E_{max}\simeq0.5$, and we see in the right-hand panel of Fig. 1 
that electron-positron pairs are created so that the particle density of the
positrons are increased in a small region around $x=0$ where the electric 
field amplitude has its maximum. Due to the ultra-strong electric field, 
the electrons are also accelerated so as to neutralize the plasma, and
hence the electric field strength decreases with time. We see in Fig. 2,
that the pair creation rate is largest initially, when the electric field
is strongest, and decreases at later times. Both the electrons and positrons
are accelerated to ultra-relativistic speeds with gamma factors of the order
100--500, and the dynamics is essentially a balance between the 
kinetic energy of the particles and the potential energy stored in the
electric field. We consider the
energy balance in Fig. 3, where we have plotted the time evolution of the
sum of the total relativistic particle energies and the electrostatic energy. 
We see that this energy is approximately conserved (solid line) when the
the full expressions for the pair creation in the continuity equations 
(\ref{ne_1D})--(\ref{np_1D}) are used. The last terms in the continuity equation account approximately
for that the electron and positron is created at some spatial difference
from each other so that the total energy of the newly created pair is 
compensated by a decrease of the electric field and hence of the electrostatic
energy. If the last terms in the right-hand sides of the continuity equations
are neglected, meaning that the pairs are created at the same point in space, 
then there is a visible increase of the total energy when the pairs are 
created (the dotted curve in Fig. 3). However, the energy loss to the pairs
is relatively small compared to the total free energy of the system, stored
initially in the electrostatic field.

In summary, we have presented a dynamical and self-consistent model
for the electron-positron pair creation by a strong electrostatic field 
in a dense plasma. The electrostatic field is excited self-consistently by
large-amplitude electron waves, which gives rise to electric fields that are
comparable with Schwinger's critical field. We have derived an approximately
energy-conserving one-dimensional model that takes into account the
decrease of the electrostatic energy as electron-positron pairs are created. 
The model presented here constitutes a first step of understanding the 
dynamics of large-amplitude electrostatic waves that are strong enough to
create electron-positron pairs. A future model should be derived from 
first principles of quantum electrodynamics, to take into account the 
distribution of the created electron positron pairs in momentum space
and the self-consistent energy conservation of the system.

\section*{Acknowledgment}

This work was partially supported by the Swedish Research Council and
by the DFG through the SFB 591.


\begin{thebibliography}{99}

\bibitem{marklund-shukla}
M. Marklund and P. K. Shukla, Rev. Mod. Phys. \textbf{78}, No. 2, in press
(2006).

\bibitem{brezin-itzykson}
E. Brezin and C. Itzykson, Phys. Rev. D \textbf{2}, 1191 (1970).

\bibitem{alkofer-etal}
R. Alkofer, M. B. Hecht, C. D. Roberts, S. M. Schmidt, and D. V. Vinnik,
Phys. Rev. Lett. \textbf{87}, 193902 (2001).

\bibitem{nitta-etal}
H. Nitta, M. Kh. Khokonov, Y. Nagata, and S. Onuki,
Phys. Rev. Lett. \textbf{93}, 180407 (2004).

\bibitem{blaschke-etal}
D. B. Blaschke, A. V. Prozorkevich, C. D. Roberts, S. M. Schmidt, and
S. A. Smolyanski,
nucl-th/0511085.

\bibitem{gies-klingmuller}
H. Gies and K. Klingm\"uller, Phys. Rev. D \textbf{72}, 065001 (2005).

\bibitem{gies-etal}
H. Gies, J. Sanchez-Guillen, R. A. V\'azquez, JHEP \textbf{08}, 067
(2005).

\bibitem{dunne-schubert}
G. V. Dunne and C. Schubert, Phys. Rev. D \textbf{72}, 105004 (2005).

\bibitem{fried-gabellini}
H. M. Fried and Y. Gabellini, Phys. Rev. D \textbf{73}, 011901(R)
(2006).

\bibitem{Schwinger}
J.\ Schwinger, Phys.\ Rev.\ \textbf{82}, 664 (1951).

\bibitem{Kajantie-Matsui}
K.\ Kajantie and T.\ Matsui, Phys.\ Lett.\ B \textbf{164}, 373 (1985).

\bibitem{Gatoff-etal}
G.\ Gatoff, A.\ K.\ Kerman, and T.\ Matsui, Phys.\ Rev.\ D
\textbf{36},114 (1987).

\bibitem{Kluger-etal}
Y.\ Kluger, J.\ M.\ Eisenberg, and B.\ Svetitsky, Phys.\ Rev.\ Lett.\
\textbf{67}, 2427 (1991).

\bibitem{Manfredi}
G.\ Manfredi, in \textit{Topics in Kinetic Theory}, Eds.\ T.\ Passot,
C.\ Sulem, and P.-L.\ Sulem, (Fields Institute Communications, American
Mathematical Society, 2005), quant-ph/0505004.

\bibitem{bulanov}
S. S. Bulanov, Phys. Rev. E \textbf{69}, 036408 (2004).

\bibitem{Bulanov-etal}
S.\ S.\ Bulanov, A.\ M. Fedotov, and F.\ Pegoraro, Phys.\ Rev.\ E
\textbf{71}, 016404 (2005).

\end{thebibliography}
\end{document}